\documentclass{article}
\usepackage{amssymb}
\usepackage{amsfonts}
\usepackage{amsmath}

\begin{document}

\begin{center}
\bigskip

{\large RENORMALIZABILITY OF A MODIFIED GENERALLY COVARIANT YANG-MILLS ACTION%
}

{\large \ }\bigskip

C. N. RAGIADAKOS

Pedagogical Institute

Mesogion 396, Agia Paraskevi, TK 15341, Greece

email: ragiadak@hol.gr, crag@pi-schools.gr

\bigskip

\textbf{ABSTRACT}
\end{center}

\begin{quote}
A modified generally covariant Yang-Mills action, which depends on the
complex structure of spacetime and not its metric, is proved to be
renormalizable. This proof makes this Lagrangian model the unique known
generally covariant four dimensional model to be renormalizable without
higher order derivatives. The first order one-loop diagrams are computed in
an appropriate gauge condition and they are found to be finite.
\end{quote}

\pagebreak

\renewcommand{\theequation}{\arabic{section}.\arabic{equation}}

\section{INTRODUCTION}

\setcounter{equation}{0}

Renormalizability seems to be the necessary criterion for a Quantum Field
Theoretic Lagrangian to be self consistent. Any physically interesting
Lagrangian has to be renormalizable in order to provide finite computations
of physical quantities. Recall that one of the cornerstones of the success
of the Standard Model was the proof of its renormalizability. But the
straightforward \textquotedblleft covariantization\textquotedblright\ of the
Standard Model action with the Einstein gravitational term is not
renormalizable. Therefore, this route has been abandoned as a possible
unification of Gravity and Quantum Field Theory. It is well known that the
main argument of the superstrings researchers is that superstrings bypass
renormalizability problem. In the present work a slightly modified generally
covariant Yang Mills action is found to be renormalizable.

The covariantized ordinary Yang Mills action
\begin{equation}
\begin{array}{l}
I_{YM}=-\frac{1}{4}\int d^{4}\!x\ \sqrt{-g}\ g^{\mu \nu }g^{\rho \sigma
}F_{\!j\mu \rho }F_{\!j\nu \sigma } \\
\\
F_{j\mu \nu }=\partial _{\mu }A_{j\nu }-\partial _{\nu }A_{j\mu
}-q\,f_{jik}A_{i\mu }A_{k\nu }%
\end{array}
\label{a1}
\end{equation}%
is invariant under the Weyl transformation, but it is not renormalizable,
because the regularization procedure generates the conformally invariant
geometric term
\begin{equation}
\begin{array}{l}
I_{W}=\int d^{4}\!x\ \sqrt{-g}\ C^{\mu \nu \rho \sigma }C_{\!\mu \nu \rho
\sigma } \\
\end{array}
\label{a2}
\end{equation}%
where $C_{\!\mu \nu \rho \sigma }$\ is the Weyl tensor. The quantization of
this action leads to inconsistencies, because its explicit dependence on
second order derivatives generates negative norm states.\ Despite the
failure to provide a self-consistent Quantum Field Theory we see that the
Weyl symmetry restricts all the permitted geometric action terms to just
one, the (\ref{a2}). The renormalizability of the Lagrangian model
considered in this work is essentially based on an extended Weyl symmetry
over the null tetrad, which does not permit even this geometric action (\ref%
{a2}).

The initial idea was an effort\cite{RAG1988} to find a four-dimensional
action which depends on the complex structure and not on the metric of the
spacetime. Recall that the two-dimensional string action has exactly this
property. Its form%
\begin{equation}
I_{S}=\frac{1}{2}\int d^{2}\!\xi \ \sqrt{-\gamma }\ \gamma ^{\alpha \beta }\
\partial _{\alpha }X^{\mu }\partial _{\beta }X^{\nu }\eta _{\mu \nu }
\label{a3}
\end{equation}%
does not essentially depend on the metric $\gamma ^{\alpha \beta }$ of the
2-dimensional surface, but it depends on its structure coordinates $(z^{0},\
z^{\widetilde{0}})$, because in these coordinates it takes the form%
\begin{equation}
I_{S}=\int d^{2}\!z\ \partial _{0}X^{\mu }\partial _{\widetilde{0}}X^{\nu
}\eta _{\mu \nu }  \label{a4}
\end{equation}%
All the wonderful properties of the string model are essentially based on
this characteristic feature of the string action.

In the case of a four dimensional manifold the structure coordinates are $%
(z^{\alpha },\ z^{\widetilde{\alpha }}),\quad \alpha =0,1$ and the complex
structure preserving transformations are $z^{\prime \alpha }=z^{\prime
\alpha }(z^{\beta }),\quad z^{\prime \widetilde{\alpha }}=z^{\prime
\widetilde{\alpha }}(z^{\widetilde{\beta }})$. The invariant action of the
model is
\begin{equation}
\begin{array}{l}
I_{G}=\int d^{4}\!z\ F_{\!j01}F_{\!j\widetilde{0}\widetilde{1}}+comp.\ conj.
\\
\\
F_{jab}=\partial _{a}A_{jb}-\partial _{a}A_{jb}-q\,f_{jik}A_{ia}A_{kb}%
\end{array}
\label{a5}
\end{equation}

Using the null tetrad one can transcribe\cite{RAG1990} this action to a the
following generally covariant form
\begin{equation}
\begin{array}{l}
I_{G}=\int d^{4}\!x\ \sqrt{-g}\ \left\{ \left( \ell ^{\mu }m^{\rho
}F_{\!j\mu \rho }\right) \left( n^{\nu }\overline{m}^{\sigma }F_{\!j\nu
\sigma }\right) +\left( \ell ^{\mu }\overline{m}^{\rho }F_{\!j\mu \rho
}\right) \left( n^{\nu }m^{\sigma }F_{\!j\nu \sigma }\right) \right\}  \\
\\
F_{j\mu \nu }=\partial _{\mu }A_{j\nu }-\partial _{\nu }A_{j\mu
}-qf_{jik}A_{i\mu }A_{k\nu }%
\end{array}
\label{a6}
\end{equation}%
where $A_{j\mu }$ is a gauge field and $(\ell _{\mu },\,n_{\mu },\,m_{\mu
},\,\overline{m}_{\mu })$ is an integrable null tetrad. The difference
between the present action and the ordinary Yang-Mills action becomes more
clear in the following form of the action.
\begin{equation}
I_{G}=-\frac{1}{8}\int d^{4}\!x\ \sqrt{-g}\ \left( 2g^{\mu \nu }\ g^{\rho
\sigma }-J^{\mu \nu }\ J^{\rho \sigma }-\overline{J^{\mu \nu }}\ \overline{%
J^{\rho \sigma }}\right) F_{\!j\mu \rho }F_{\!j\nu \sigma }  \label{a7}
\end{equation}%
where $g_{\mu \nu }$ is the metric and $J_{\mu }^{\;\nu }$ is the tensor of
the integrable complex structure derived from the null tetrad\cite{FLAHE1974}
using the following relations

\begin{equation}
\begin{array}{l}
g_{\mu \nu }=\ell _{\mu }n_{\nu }+n_{\mu }\ell _{\nu }-m_{{}\mu }\overline{m}%
_{\nu }-\overline{m}_{\mu }m_{\nu } \\
\\
J_{\mu }^{\;\nu }=i(\ell _{\mu }n^{\nu }-n_{\mu }\ell ^{\nu }-m_{\mu }%
\overline{m}^{\nu }+\overline{m}_{\mu }m^{\nu })%
\end{array}
\label{a8}
\end{equation}%
The integrability condition of the complex structure implies the Frobenius
integrability conditions of the pairs $(\ell _{\mu },\,\,m_{\mu })$ and $%
(n_{\mu },\,\overline{m}_{\mu })$. That is

\begin{equation}
\begin{array}{l}
(\ell ^{\mu }m^{\nu }-\ell ^{\nu }m^{\mu })(\partial _{\mu }\ell _{\nu
})=0\;\;\;\;,\;\;\;\;(\ell ^{\mu }m^{\nu }-\ell ^{\nu }m^{\mu })(\partial
_{\mu }m_{\nu })=0 \\
\\
(n^{\mu }m^{\nu }-n^{\nu }m^{\mu })(\partial _{\mu }n_{\nu
})=0\;\;\;\;,\;\;\;\;(n^{\mu }m^{\nu }-n^{\nu }m^{\mu })(\partial _{\mu
}m_{\nu })=0%
\end{array}
\label{a9}
\end{equation}%
Frobenius theorem states that there are four complex functions $(z^{\alpha
},\;z^{\widetilde{\alpha }}),$\ $\alpha =0,\ 1$ , such that

\begin{equation}
dz^{\alpha }=f_{\alpha }\ \ell _{\mu }dx^{\mu }+h_{\alpha }\ m_{\mu }dx^{\mu
}\;\;\;\;,\;\;\;dz^{\widetilde{\alpha }}=f_{\widetilde{\alpha }}\ n_{\mu
}dx^{\mu }+h_{\widetilde{\alpha }}\ \overline{m}_{\mu }dx^{\mu }\;
\label{a10}
\end{equation}%
These four functions are the structure coordinates of the (integrable)
complex structure used in (\ref{a5}). In the present case of Lorentzian
spacetimes the coordinates $z^{\widetilde{\alpha }}$ are not complex
conjugate of $z^{\alpha }$, because $J_{\mu }^{\;\nu }$ is no longer a real
tensor\cite{FLAHE1976}. This peculiar property\cite{RAG1991} was used by the
author\cite{RAG1999} to show that the particle spectrum of the present
Lagrangian model is very rich, while the static potential of a source is no
longer $\frac{1}{r}$ but it is linear.

A typical example of four dimensional complex structure compatible with the
Minkowski metric are the light-cone coordinates determined by the following
null tetrad%
\begin{equation}
\begin{array}{l}
E_{\mu }^{0}\equiv L_{\mu }=\frac{1}{\sqrt{2}}(1,\ -1,\ 0,\ 0) \\
E_{\mu }^{\widetilde{0}}\equiv N_{\mu }=\frac{1}{\sqrt{2}}(1,\ 1,\ 0,\ 0) \\
E_{\mu }^{1}\equiv M_{\mu }=\frac{1}{\sqrt{2}}(0,\ 0,\ 1,\ i) \\
E_{\mu }^{\widetilde{1}}\equiv \overline{M}_{\mu }=\frac{1}{\sqrt{2}}(0,\
0,\ 1,\ -i)%
\end{array}
\label{a11}
\end{equation}%
which will be used in the present work. The general null tetrad will be
expanded around this simple form.

In the case of the two-dimensional string action, no integrability
conditions are required because any orientable two dimensional manifold is a
complex manifold. But in the present case the conditions (\ref{a9}) must be
introduced in the action using the Lagrange multiplier form
\begin{equation}
\begin{array}{l}
I_{C}=-\int d^{4}\!x\ \{\phi _{0}(\ell ^{\mu }m^{\nu }-\ell ^{\nu }m^{\mu
})(\partial _{\mu }\ell _{\nu })+ \\
\\
\qquad +\phi _{1}(\ell ^{\mu }m^{\nu }-\ell ^{\nu }m^{\mu })(\partial _{\mu
}m_{\nu })+\phi _{\widetilde{0}}(n^{\mu }\overline{m}^{\nu }-n^{\nu }%
\overline{m}^{\mu })(\partial _{\mu }n_{\nu })+ \\
\\
\qquad +\phi _{\widetilde{1}}(n^{\mu }\overline{m}^{\nu }-n^{\nu }\overline{m%
}^{\mu })(\partial _{\mu }\overline{m}_{\nu })+c.conj.\}%
\end{array}
\label{a12}
\end{equation}%
The complete action $I=I_{G}+I_{C}$ is self-consistent and it was quantized
using the canonical (Dirac) quantization technique\cite{RAG1990} and the
path integral (BRST) technique\cite{RAG1992}.

The local symmetries of the action are a) the well known local gauge
transformations, b) the reparametrization symmetry as it is the case in any
generally covariant action and c) the following extended Weyl transformation
of the tetrad
\begin{equation}
\begin{array}{l}
\ell _{\mu }^{\prime }=\chi _{1}\ell _{\mu }\quad ,\quad n_{\mu }^{\prime
}=\chi _{2}n_{\mu }\quad ,\quad m_{\mu }^{\prime }=\chi m_{\mu } \\
\\
\phi _{0}^{\prime }=\phi _{0}\frac{\chi _{2}\overline{\chi }}{\chi _{1}}%
\quad ,\quad \phi _{1}^{\prime }=\phi _{1}\frac{\chi _{2}\overline{\chi }}{%
\chi } \\
\\
\phi _{\widetilde{0}}^{\prime }=\phi _{\widetilde{0}}\frac{\chi _{1}\chi }{%
\chi _{2}}\quad ,\quad \phi _{\widetilde{1}}^{\prime }=\phi _{\widetilde{1}}%
\frac{\chi _{1}\chi }{\overline{\chi }} \\
\\
g^{\prime }=g(\chi _{1}\chi _{2}\chi \overline{\chi })^{2}%
\end{array}
\label{a13}
\end{equation}%
where $\chi _{1},\chi _{2}$ are real functions and $\chi $ is a complex one.

\section{GAUGE FIELD PROPAGATOR IN THE LANDAU AND FEYNMAN GAUGES}

\setcounter{equation}{0}

The enhanced conformal symmetry makes the present action unique and as far
as I know the field propagator has never been considered in the literature.
Therefore the gauge field propagator will be computed in the well known
Landau and Feynman gauges, in order to familiarize the reader with the
peculiarities of the present gauge field action. A more appropriate gauge
field condition will be used in the present work and the path integral
(BRST) quantization\cite{RAG1992} in this gauge will be described in the
next sections. As usual the Feynman and Landau gauges are introduced in the
path integral quantization through a term $\frac{1}{\alpha }(\eta ^{\mu \nu
}\partial _{\mu }A_{j\nu })^{2}$ in the effective action. The choices $a=1$
or $\alpha =0$\ are referred as Feynman and Landau gauges respectively.
Following the well known path integral technique, the gauge field propagator
(for arbitrary $\alpha $) is%
\begin{equation}
\langle TA_{i\mu }(x)A_{j\nu }(y)\rangle =-i\delta _{ij}\int \frac{d^{4}k}{%
(2\pi )^{4}}e^{ik(y-x)}\Delta _{\mu \nu }(k)  \label{b1}
\end{equation}%
where $\Delta _{\mu \nu }(k)$ satisfies the relation%
\begin{equation}
\begin{array}{l}
\lbrack (L^{\rho }M^{\mu }-L^{\mu }M^{\rho })(N^{\lambda }\overline{M}^{\nu
}-N^{\nu }\overline{M}^{\lambda })+(N^{\rho }\overline{M}^{\mu }-N^{\mu }%
\overline{M}^{\rho })(L^{\lambda }M^{\nu }-L^{\nu }M^{\lambda })+ \\
+(L^{\rho }\overline{M}^{\mu }-L^{\mu }\overline{M}^{\rho })(N^{\lambda
}M^{\nu }-N^{\nu }M^{\lambda })+(N^{\rho }M^{\mu }-N^{\mu }M^{\rho
})(L^{\lambda }\overline{M}^{\nu }-L^{\nu }\overline{M}^{\lambda })- \\
-\frac{1}{\alpha }\eta ^{\rho \mu }\eta ^{\lambda \nu }]k_{\rho }k_{\lambda
}\Delta _{\mu \nu }(k)=-\delta _{\sigma }^{\mu }%
\end{array}
\label{b2}
\end{equation}%
which is found after the expansion of the action around the light-cone
(integrable) null tetrad (\ref{a11}). Throughout this work the general null
tetrad will be expanded around the light-cone one, because the calculations
are highly simplified.

Expanding $\Delta _{\nu \sigma }(k)$ in this null tetrad
\begin{equation}
\begin{array}{l}
\Delta _{\nu \sigma }=H_{00}L_{\nu }L_{\sigma }+H_{01}(L_{\nu }N_{\sigma
}+L_{\sigma }N_{\nu })+H_{02}(L_{\nu }M_{\sigma }+L_{\sigma }M_{\nu })+ \\
\qquad +\overline{H}_{02}(L_{\nu }\overline{M}_{\sigma }+L_{\sigma }%
\overline{M}_{\nu })+H_{11}N_{\nu }N_{\sigma }+H_{12}(N_{\nu }M_{\sigma
}+N_{\sigma }M_{\nu })+ \\
\qquad +\overline{H}_{12}(N_{\nu }\overline{M}_{\sigma }+N_{\sigma }%
\overline{M}_{\nu })+H_{22}M_{\nu }M_{\sigma }+ \\
\qquad +H_{23}(M_{\nu }\overline{M}_{\sigma }+M_{\sigma }\overline{M}_{\nu
})+\overline{H}_{22}\overline{M}_{\nu }\overline{M}_{\sigma }%
\end{array}
\label{b3}
\end{equation}%
and substituting into the above relation (\ref{b2}), a system of linear
equations is derived, which can be directly solved. The final result is
\begin{equation}
\begin{array}{l}
H_{00}=\frac{(Nk)(Nk)}{2(Mk)(\overline{M}k)k^{2}}+\frac{(\alpha -1)(Nk)(Nk)}{%
k^{4}} \\
H_{01}=\frac{1}{k^{2}}\left[ 1-\frac{(Lk)(Nk)}{2(Mk)(\overline{M}k)k^{2}}+%
\frac{(\alpha -1)(Lk)(Nk)}{k^{2}}\right] \\
H_{02}=\frac{(1-\alpha )(Nk)(\overline{M}k)}{k^{4}} \\
H_{11}=\frac{(Lk)(Lk)}{2(Mk)(\overline{M}k)k^{2}}+\frac{(\alpha -1)(Lk)(Lk)}{%
k^{4}} \\
H_{12}=\frac{(1-\alpha )(Lk)(\overline{M}k)}{k^{4}} \\
H_{22}=-\frac{(\overline{M}k)(\overline{M}k)}{2(Lk)(Nk)k^{2}}+\frac{(\alpha
-1)(\overline{M}k)(\overline{M}k)}{k^{4}} \\
H_{23}=\frac{1}{k^{2}}\left[ -1+\frac{(Mk)(\overline{M}k)}{2(Lk)(Nk)}+\frac{%
(\alpha -1)(Mk)(\overline{M}k)}{k^{4}}\right]%
\end{array}
\label{b4}
\end{equation}%
where the short notation $(E_{a}k)\equiv E_{a}^{\mu }k_{\mu }$\ is used.\ In
fact these are the light-cone coordinates of the four-vector $k_{\mu }$.
This short light-cone notation will be used throughout this work in order to
keep track of the initial tetrad structure of the different Lagrangian terms..

In the Landau gauge ($\alpha =0$) the Fourier transform of the gauge field
propagator takes the form%
\begin{equation}
\begin{array}{l}
\langle TA_{i\mu }(x)A_{j\nu }(y)\rangle _{F}=-\frac{i\delta _{ij}}{k^{2}}%
[\eta _{\mu \nu }-\frac{k_{\mu }k_{\nu }}{k^{2}}+\frac{(Nk)(Nk)}{2(Mk)(%
\overline{M}k)}L_{\mu }L_{\nu }+ \\
\qquad +\frac{(Lk)(Lk)}{2(Mk)(\overline{M}k)}N_{\mu }N_{\nu }-\frac{(Lk)(Nk)%
}{2(Mk)(\overline{M}k)}(L_{\mu }N_{\nu }+L_{\nu }N_{\mu })-\frac{(\overline{M%
}k)(\overline{M}k)}{2(Lk)(Nk)}M_{\mu }M_{\nu }+ \\
\qquad +\frac{(Mk)(\overline{M}k)}{2(Lk)(Nk)}(M_{\mu }\overline{M}_{\nu
}+M_{\nu }\overline{M}_{\mu })-\frac{(Mk)(Mk)}{2(Lk)(Nk)}\overline{M}_{\mu }%
\overline{M}_{\nu }]%
\end{array}
\label{b5}
\end{equation}%
Notice that in addition to the ordinary term $\eta _{\mu \nu }-\frac{k_{\mu
}k_{\nu }}{k^{2}}$\ of the gauge field propagator it contains
non-conventional terms too. The difference of the present gauge field action
(\ref{a7}) from the ordinary one appears in the gauge field propagator too,
as we should expect. In the Feynman gauge ($\alpha =1$) only the ordinary
part of the propagator changes to the well known form. The additional
non-conventional terms remain the same.

\section{AN APPROPRIATE GAUGE CONDITION}

\setcounter{equation}{0}

In the Landau and Feynman gauges, the gauge field propagators are very
complicated. Therefore they are not convenient for the computation of the
Feynman diagrams. It was found that the most convenient gauge condition is
\begin{equation}
M^{\mu }\partial _{\mu }(\overline{M}A_{j})+\overline{M}^{\mu }\partial
_{\mu }(MA_{j})=0  \label{c1}
\end{equation}%
where $(E^{a}A_{j})\equiv E^{a\mu }A_{j\mu }$ are the light-cone coordinates
of the gauge field $A_{j\mu }$. In section 7 explicit calculations of the
first order one-loop diagrams will be performed. The great advantage of this
precise gauge condition is that these diagrams are found to be finite. That
is no counterterms appear!

In the path integral formulation, the validity of a gauge condition is
formally assured through the non-annihilation of the Faddeev-Popov
determinant. It will be checked below in the case of an Abelian $U(1)$\
gauge field. It is generally assumed that the same results are
perturbatively extended to the non-Abelian cases modulo possible Gribov
ambiguities. The above gauge condition yields the following Faddeev-Popov
operator%
\begin{equation}
M_{FP}=-\left( \frac{\partial ^{2}}{\partial y^{2}}+\frac{\partial ^{2}}{%
\partial z^{2}}\right) \cdot   \label{c2}
\end{equation}

The determinant of this operator does not vanish, because it has no regular
asymptotically vanishing eigenfunction with zero eigenvalue. One can see it
by simply writing this operator in polar coordinates and making a Fourier
expansion. Then we see that the zero modes must satisfy the following
differential equation%
\begin{equation}
\left( \frac{\partial ^{2}}{\partial \rho ^{2}}+\frac{1}{\rho }\frac{%
\partial }{\partial \rho }-\frac{n^{2}}{\rho ^{2}}\right) \Lambda
_{n}(t,x,\rho )=0  \label{c3}
\end{equation}%
For $n\neq 0$ the general solution of this equation is%
\begin{equation}
\Lambda _{n}(t,x,\rho )=h_{1n}(t,x)\rho ^{n}+h_{2n}(t,x)\rho ^{-n}
\label{c4}
\end{equation}%
which is regular at $\rho =0$\ if $h_{2}=0$ and it vanishes at infinity if $%
h_{1}=0$. For $n=0$ the solution is
\begin{equation}
\Lambda _{0}(t,x,\rho )=h_{10}(t,x)+h_{20}(t,x)\ln \rho  \label{c5}
\end{equation}%
which does not satisfy the regularity conditions. Hence we see that the
kernel of the Faddeev-Popov operator contains only the zero function.

One should not be confused by the apparent permitted gauge transformation%
\begin{equation}
A_{\mu }^{\prime }=A_{\mu }-\partial _{\mu }\Lambda (t,x)  \label{c6}
\end{equation}%
because the asymptotic annihilation is assumed in all space directions. $%
\Lambda (t,x)$ must vanish because at $\rho $-infinity it is the same
function. Recall that the same argument is applied to the case of the axial
gauge condition of the electromagnetic field too.

In the conventional procedure, the non-vanishing of the Faddeev-Popov
determinant means that the gauge condition uniquely fixes the gauge freedom
of the action. The additional point, one should clarify, is that the precise
gauge can always be reached starting from any regular asymptotically
vanishing field configuration $A_{\mu }(x)$. One can see that it is
reachable, if there is a regular asymptotically vanishing solution of the
differential equation%
\begin{equation}
\left( \frac{\partial ^{2}}{\partial y^{2}}+\frac{\partial ^{2}}{\partial
z^{2}}\right) \Lambda =M^{\mu }\partial _{\mu }(\overline{M}A_{j})+\overline{%
M}^{\mu }\partial _{\mu }(MA_{j})\equiv f(x)  \label{c7}
\end{equation}

In polar coordinates and after a Fourier expansion it becomes the following
ordinary differential equation%
\begin{equation}
\left( \frac{\partial ^{2}}{\partial \rho ^{2}}+\frac{1}{\rho }\frac{%
\partial }{\partial \rho }-\frac{n^{2}}{\rho ^{2}}\right) \Lambda
_{n}(t,x,\rho )=f_{n}(t,x,\rho )  \label{c8}
\end{equation}%
which always admits a solution with initial conditions%
\begin{equation}
\Lambda _{n}(t,x,0)=0\quad ,\quad \frac{d\Lambda _{n}}{d\rho }(t,x,0)=0
\label{c9}
\end{equation}%
The above analysis of the convenient gauge condition shows that it is well
defined and it may be used to determine the gauge field propagator.

\section{LAGRANGIAN EXPANSION AND PROPAGATORS}

\setcounter{equation}{0}

The Dirac\cite{RAG1990} and BRST\cite{RAG1992} quantizations of the model
will be used to study its renormalizability. The path-integral (BRST)
quantization can be accomplished by simply following the ordinary steps. We
first see that the local symmetries of the complete action are the usual
gauge symmetry, reparametrization and the extended Weyl transformations. For
every local symmetry we have to assume a gauge condition. Here we must be
careful to impose convenient gauge conditions such that the induced
Faddeev-Popov determinant to have vanishing the upper diagonal elements in
order to be reduced down into the product of the three determinants which
correspond to the three local symmetries of the action. The gauge symmetry
is fixed using the appropriate gauge condition (\ref{c1}). The additional
extended Weyl symmetry of the tetrad is fixed using the following conditions%
\begin{equation}
\begin{array}{l}
\ell ^{\mu }N_{\mu }-1=0\quad ,\quad n^{\mu }L_{\mu }-1=0 \\
\overline{m}^{\mu }M_{\mu }+1=0\quad ,\quad m^{\mu }\overline{M}_{\mu }+1=0
\\
\end{array}
\label{d1}
\end{equation}%
The convenient conditions which fix the reparametrization symmetry are%
\begin{equation}
\begin{array}{l}
L^{\mu }\ell _{\mu }n^{\nu }L_{\nu }=0\quad ,\quad N^{\mu }n_{\mu }\ell
^{\nu }N_{\nu }=0 \\
M^{\mu }m_{\mu }\overline{m}^{\nu }M_{\nu }=0\quad ,\quad \overline{M}^{\mu }%
\overline{m}_{\mu }m^{\nu }\overline{M}_{\nu }=0 \\
\end{array}
\label{d2}
\end{equation}%
Then the Faddeev-Popov terms of the effective Lagrangian are the following%
\cite{RAG1992}%
\begin{equation}
\begin{array}{l}
I_{FP}=\int d^{4}x\{-\frac{1}{2\alpha }[M^{\mu }\partial _{\mu }(\overline{M}%
A_{j})+\overline{M}^{\mu }\partial _{\mu }(MA_{j})]^{2}+B_{1}(\ell ^{\mu
}N_{\mu }-1)+ \\
\qquad +B_{2}(n^{\mu }L_{\mu }-1)+B_{3}(\overline{m}^{\mu }M_{\mu
}+1)+B_{4}(m^{\mu }\overline{M}_{\mu }+1)+ \\
\qquad +B_{5}(L^{\mu }\ell _{\mu })(n^{\nu }L_{\nu })+B_{6}(N^{\mu }n_{\mu
})(\ell ^{\nu }N_{\nu })+ \\
\qquad +B_{7}(M^{\mu }m_{\mu })(\overline{m}^{\nu }M_{\nu })+B_{8}(\overline{%
M}^{\mu }\overline{m}_{\mu })(m^{\nu }\overline{M}_{\nu })+ \\
\qquad +M^{\mu }(\partial _{\mu }\overline{d}_{j})[\overline{M}^{\nu
}(\partial _{\nu }d_{j})-qf_{jik}d_{i}(\overline{M}^{\nu }A_{k\nu })] \\
\qquad +\overline{M}^{\mu }(\partial _{\mu }\overline{d}_{j})[M^{\nu
}(\partial _{\nu }d_{j})-qf_{jik}d_{i}(M^{\nu }A_{k\nu })] \\
\qquad -\overline{c}_{1}L^{\mu }[c^{\nu }(\partial _{\nu }\ell _{\mu })+\ell
_{\nu }(\partial _{\mu }c^{\nu })]-\overline{c}_{2}N^{\mu }[c^{\nu
}(\partial _{\nu }n_{\mu })+n_{\nu }(\partial _{\mu }c^{\nu })]- \\
\qquad -\overline{c}_{3}M^{\mu }[c^{\nu }(\partial _{\nu }m_{\mu })+m_{\nu
}(\partial _{\mu }c^{\nu })]-\overline{c}_{4}\overline{M}^{\mu }[c^{\nu
}(\partial _{\nu }\overline{m}_{\mu })+\overline{m}_{\nu }(\partial _{\mu
}c^{\nu })]\} \\
\end{array}
\label{d3}
\end{equation}%
where $\overline{d}_{j}$\ and $d_{j}$\ are the ghost fields which correspond
to the gauge field condition and $\overline{c}_{i}\ ,\ c_{i}$ are the ghost
fields which correspond to the reparametrization symmetry. The extended Weyl
symmetry on the tetrad does not generate any ghost field.

In order to compute the Feynman diagrams we have first to expand the
Lagrangian around a classical solution of the field equations. In the
present case it is convenient to expand the general null tetrad around the
trivial light-cone tetrad $E_{a}^{\mu }$ that we have chosen to introduce
the conditions which fix the reparametrization and Weyl symmetries. That is,
we consider the expansion%
\begin{equation}
\begin{array}{l}
\ell ^{\mu }=L^{\mu }+\gamma \varepsilon _{\widetilde{0}}^{\mu } \\
n^{\mu }=N^{\mu }+\gamma \varepsilon _{0}^{\mu } \\
m^{\mu }=M^{\mu }-\gamma \varepsilon _{\widetilde{1}}^{\mu }%
\end{array}
\label{d4}
\end{equation}%
where $\gamma $\ is a dimensionless constant. Notice that in the Lagrangian
there is no dimensional constant, which could generate non-renormalizable
counterterms through the regularization procedure. In this tetrad expansion,
the conditions become%
\begin{equation}
\begin{array}{l}
\varepsilon _{\widetilde{0}}^{\mu }N_{\mu }=0\quad ,\quad \varepsilon
_{0}^{\mu }L_{\mu }=0\quad ,\quad \varepsilon _{1}^{\mu }M_{\mu }=0 \\
\varepsilon _{\widetilde{0}}^{\mu }L_{\mu }-\gamma \lbrack (\varepsilon _{%
\widetilde{0}}^{\nu }M_{\nu })(\varepsilon _{1}^{\rho }L_{\rho
})+(\varepsilon _{\widetilde{0}}^{\nu }\overline{M}_{\nu })(\varepsilon _{%
\widetilde{1}}^{\rho }L_{\rho })]+O(\gamma ^{2})=0 \\
\varepsilon _{0}^{\mu }N_{\mu }-\gamma \lbrack (\varepsilon _{0}^{\nu
}M_{\nu })(\varepsilon _{1}^{\rho }N_{\rho })+(\varepsilon _{0}^{\nu }%
\overline{M}_{\nu })(\varepsilon _{\widetilde{1}}^{\rho }N_{\rho
})]+O(\gamma ^{2})=0 \\
\varepsilon _{1}^{\mu }\overline{M}_{\mu }-\gamma \lbrack (\varepsilon
_{1}^{\nu }L_{\nu })(\varepsilon _{0}^{\rho }\overline{M}_{\rho
})+(\varepsilon _{1}^{\nu }N_{\nu })(\varepsilon _{\widetilde{0}}^{\rho }%
\overline{M}_{\rho })]+O(\gamma ^{2})=0 \\
\end{array}
\label{d5}
\end{equation}

They can be solved and replaced back into the action, which is so expanded
in the dimensionless coupling constants $\gamma $ and $q$. The first terms
of this expansion of the $I_{G}$ part of the action are the following%
\begin{equation}
\begin{array}{l}
I_{G}\simeq \int d^{4}x\{[(LM\partial A_{j})(N\overline{M}\partial A_{j})+(L%
\overline{M}\partial A_{j})(NM\partial A_{j})]- \\
\qquad -qf_{jik}[(LA_{i})(MA_{k})(N\overline{M}\partial A_{j})+(NA_{i})(%
\overline{M}A_{k})(LM\partial A_{j})+c.c]+ \\
\qquad +\gamma \lbrack (M\varepsilon _{\widetilde{0}})(M\overline{M}\partial
A_{j})(N\overline{M}\partial A_{j})-(L\varepsilon _{\widetilde{1}%
})(LN\partial A_{j})(N\overline{M}\partial A_{j})+ \\
\qquad +(N\varepsilon _{1})(LM\partial A_{j})(LN\partial A_{j})-(\overline{M}%
\varepsilon _{0})(LM\partial A_{j})(M\overline{M}\partial A_{j})+c.c]+ \\
\qquad +q^{2}f_{jik}f_{ji^{\prime }k^{\prime
}}[(LA_{i})(MA_{k})(NA_{i^{\prime }})(\overline{M}A_{k^{\prime }})+c.c]\} \\
\end{array}
\label{d6}
\end{equation}%
where short notations of the form $(LM\partial A_{j})=(L^{\mu }M^{\nu
}-L^{\nu }M^{\mu })(\partial _{\mu }A_{j\nu })$ etc are used in order to
simplify the appearance of this and the following expressions. The first
terms of the $I_{C}$ part of the action are

\begin{equation}
\begin{array}{l}
I_{C}\simeq \int d^{4}x\{-[\phi _{0}L^{\nu }\partial _{\nu }(L\varepsilon _{%
\widetilde{1}})+\phi _{1}M^{\nu }\partial _{\nu }(M\varepsilon _{\widetilde{0%
}})+ \\
\qquad +\phi _{\widetilde{0}}N^{\nu }\partial _{\nu }(N\varepsilon
_{1})+\phi _{\widetilde{1}}\overline{M}^{\nu }\partial _{\nu }(\overline{M}%
\varepsilon _{0})+c.c.]- \\
\qquad -\gamma \lbrack \phi _{0}(M\varepsilon _{\widetilde{0}})[M^{\nu
}\partial _{\nu }(L\varepsilon _{1})-\overline{M}^{\nu }\partial _{\nu
}(L\varepsilon _{\widetilde{1}})]+ \\
\qquad +\phi _{1}(L\varepsilon _{\widetilde{1}})[L^{\nu }\partial _{\nu
}(M\varepsilon _{0})-N^{\nu }\partial _{\nu }(M\varepsilon _{\widetilde{0}%
})]+ \\
\qquad +\phi _{\widetilde{0}}(\overline{M}\varepsilon _{0})[\overline{M}%
^{\nu }\partial _{\nu }(N\varepsilon _{\widetilde{1}})-M^{\nu }\partial
_{\nu }(N\varepsilon _{1})]+ \\
\qquad +\phi _{\widetilde{1}}(N\varepsilon _{1})[N^{\nu }\partial _{\nu }(%
\overline{M}\varepsilon _{\widetilde{0}})-L^{\nu }\partial _{\nu }(\overline{%
M}\varepsilon _{0})]+c.c.]\} \\
\end{array}
\label{d7}
\end{equation}%
The first terms of the $I_{FP}$ part of the action are%
\begin{equation}
\begin{array}{l}
I_{FP}\simeq \int d^{4}x\{-\frac{1}{2\alpha }[M^{\mu }\partial _{\mu }(%
\overline{M}A_{j})+\overline{M}^{\mu }\partial _{\mu }(MA_{j})]^{2}]- \\
\qquad -2\overline{d}_{j}M^{\mu }\overline{M}^{\nu }(\partial _{\mu
}\partial _{\nu }d_{j})-\overline{c}_{1}L^{\mu }\partial _{\mu }(Lc)- \\
\qquad -\overline{c}_{2}N^{\mu }\partial _{\mu }(Nc)-\overline{c}_{3}M^{\mu
}\partial _{\mu }(Mc)-\overline{c}_{4}\overline{M}^{\mu }\partial _{\mu }(%
\overline{M}c)- \\
\qquad -qf_{jik}[M^{\mu }(\partial _{\mu }\overline{d}_{j})d_{i}(\overline{M}%
A_{k})+\overline{M}^{\mu }(\partial _{\mu }\overline{d}_{j})d_{i}(MA_{k})]+
\\
\qquad +\gamma \lbrack \overline{c}_{1}(L\varepsilon _{1})L^{\mu }\partial
_{\mu }(Mc)+\overline{c}_{1}(L\varepsilon _{\widetilde{1}})L^{\mu }\partial
_{\mu }(\overline{M}c)+ \\
\qquad +\overline{c}_{2}(N\varepsilon _{1})N^{\mu }\partial _{\mu }(Mc)+%
\overline{c}_{2}(N\varepsilon _{\widetilde{1}})N^{\mu }\partial _{\mu }(%
\overline{M}c)+ \\
\qquad +\overline{c}_{3}(M\varepsilon _{0})N^{\mu }\partial _{\mu }(Lc)+%
\overline{c}_{3}(M\varepsilon _{\widetilde{0}})M^{\mu }\partial _{\mu }(Nc)+
\\
\qquad +\overline{c}_{4}(\overline{M}\varepsilon _{0})\overline{M}^{\mu
}\partial _{\mu }(Lc)+\overline{c}_{4}(\overline{M}\varepsilon _{\widetilde{0%
}})\overline{M}^{\mu }\partial _{\mu }(Nc)]\} \\
\end{array}
\label{d8}
\end{equation}%
where the already defined short light-cone notation is used.

The zeroth order terms of this action expansion determine the field
propagators. The Fourier transforms of the gauge field propagator has the
following for general $\alpha $%
\begin{equation}
\begin{array}{l}
\langle TA_{i\mu }(x)A_{j\nu }(y)\rangle _{F}=-\frac{i\delta _{ij}}{4(Mk)(%
\overline{M}k)}[\frac{\alpha (Nk)(Nk)}{(Mk)(\overline{M}k)}L_{\mu }L_{\nu }+%
\frac{\alpha (Lk)(Lk)}{(Mk)(\overline{M}k)}N_{\mu }N_{\nu }+ \\
\qquad +\frac{\alpha (Lk)(Nk)-2(Mk)(\overline{M}k)}{(Mk)(\overline{M}k)}%
(L_{\mu }N_{\nu }+L_{\nu }N_{\mu })-\frac{\alpha (Nk)(\overline{M}k)}{(Mk)(%
\overline{M}k)}(L_{\mu }M_{\nu }+L_{\nu }M_{\mu })- \\
\qquad -\frac{\alpha (Nk)(Mk)}{(Mk)(\overline{M}k)}(L_{\mu }\overline{M}%
_{\nu }+L_{\nu }\overline{M}_{\mu })-\frac{\alpha (Lk)(\overline{M}k)}{(Mk)(%
\overline{M}k)}(N_{\mu }M_{\nu }+N_{\nu }M_{\mu })- \\
\qquad -\frac{\alpha (Lk)(Mk)}{(Mk)(\overline{M}k)}(L_{\mu }\overline{M}%
_{\nu }+L_{\nu }\overline{M}_{\mu })+\frac{(\overline{M}k)(\overline{M}%
k)(\alpha (Lk)(Nk)+(Mk)(\overline{M}k))}{(Lk)(Nk)(Mk)(\overline{M}k)}M_{\mu
}M_{\nu }- \\
\qquad +\frac{\alpha (Lk)(Nk)-(Mk)(\overline{M}k)}{(Lk)(Nk)}(M_{\mu }%
\overline{M}_{\nu }+M_{\nu }\overline{M}_{\mu })+\frac{(Mk)(Mk)(\alpha
(Lk)(Nk)+(Mk)(\overline{M}k))}{(Lk)(Nk)(Mk)(\overline{M}k)}\overline{M}_{\mu
}\overline{M}_{\nu }] \\
\end{array}
\label{d9}
\end{equation}%
One can easily find that in the special gauge $\alpha =0$ the non-vanishing
terms of gauge field propagator are
\begin{equation}
\begin{array}{l}
\langle T(LA_{i})(NA_{j})\rangle _{F}=\frac{i\delta _{ij}}{2(Mk)(\overline{M}%
k)} \\
\langle T(MA_{i})(MA_{j})\rangle _{F}=-\frac{i(Mk)(Mk)\delta _{ij}}{%
4(Lk)(Nk)(Mk)(\overline{M}k)} \\
\langle T(\overline{M}A_{i})(\overline{M}A_{j})\rangle _{F}=-\frac{i(%
\overline{M}k)(\overline{M}k)\delta _{ij}}{4(Lk)(Nk)(Mk)(\overline{M}k)} \\
\langle T(MA_{i})(\overline{M}A_{j})\rangle _{F}=\frac{i\delta _{ij}}{%
4(Lk)(Nk)} \\
\end{array}
\label{d10}
\end{equation}%
where the previously defined light-cone short notation is used%
\begin{equation}
\begin{array}{l}
(Lk)=\frac{k^{0}-k^{1}}{\sqrt{2}} \\
(Nk)=\frac{k^{0}+k^{1}}{\sqrt{2}} \\
(Mk)=\frac{k^{2}+ik^{3}}{\sqrt{2}} \\
\end{array}
\label{d11}
\end{equation}

Notice that this propagator is essentially the product of two well known
2-dimensional scalar field propagator%
\begin{equation}
D_{L(E)}=\int \frac{d^{2}k}{(2\pi )^{2}}\frac{e^{ikx}}{k^{2}+i\varepsilon }=%
\frac{i}{4\pi }\int \frac{dt}{t}e^{-i(x^{2}-i\varepsilon )t}  \label{d12}
\end{equation}%
where the indices $L$\ and $E$\ correspond to the signatures $(+,-)$ and $%
(-,-)$ respectively. This propagator is logarithmically divergent, but the
difference $D(x)-D(x_{0})$\ is apparently finite. One can easily find that
the explicit form of the present gauge field propagator is%
\begin{equation}
\begin{array}{l}
\langle T\left( LA_{i}(0)\right) \left( NA_{j}(x)\right) \rangle =-i\delta
_{ij}\delta (x^{0})\delta (x^{1})D_{E}(x^{2},x^{3}) \\
\langle T\left( MA_{i}(0)\right) \left( MA_{j}(x)\right) \rangle =i\delta
_{ij}D_{L}(x^{0},x^{1})M^{\mu }M^{\nu }\partial _{\mu }\partial _{\nu
}D_{E}(x^{2},x^{3}) \\
\langle T\left( \overline{M}A_{i}(0)\right) \left( \overline{M}%
A_{j}(x)\right) \rangle =i\delta _{ij}D_{L}(x^{0},x^{1})\overline{M}^{\mu }%
\overline{M}^{\nu }\partial _{\mu }\partial _{\nu }D_{E}(x^{2},x^{3}) \\
\langle T\left( MA_{i}(0)\right) \left( \overline{M}A_{j}(x)\right) \rangle
=i\delta _{ij}D_{L}(x^{0},x^{1})\delta (x^{2})\delta (x^{3}) \\
\end{array}
\label{d13}
\end{equation}

The Fourier transforms of the other field propagators are%
\begin{equation}
\begin{array}{l}
\langle T\phi _{0}(L\varepsilon _{\widetilde{1}})\rangle _{F}=-\frac{1}{(Lk)}%
\quad ,\quad \langle T\phi _{1}(M\varepsilon _{\widetilde{0}})\rangle _{F}=-%
\frac{1}{(Mk)} \\
\langle T\phi _{\widetilde{0}}(N\varepsilon _{1})\rangle _{F}=-\frac{1}{(Nk)}%
\quad ,\quad \langle T\phi _{\widetilde{1}}(\overline{M}\varepsilon
_{0})\rangle _{F}=-\frac{1}{(\overline{M}k)} \\
\langle T\overline{c}_{1}(Lc)\rangle _{F}=\frac{1}{(Lk)}\quad ,\quad \langle
T\overline{c}_{2}(Nc)\rangle _{F}=\frac{1}{(Nk)} \\
\langle T\overline{c}_{3}(Mc)\rangle _{F}=\frac{1}{(Mk)}\quad ,\quad \langle
T\overline{c}_{4}(\overline{M}c)\rangle _{F}=\frac{1}{(\overline{M}k)} \\
\langle Td_{i}\overline{d}_{j}\rangle _{F}=\frac{i\delta _{ij}}{2(Mk)(%
\overline{M}k)} \\
\end{array}
\label{d14}
\end{equation}

Notice that there is no tetrad-tetrad propagator. Only $\phi _{b}-$tetrad
propagators exist. This implies that tere is no loop diagram with $\phi _{b}$%
\ external lines. The one-particle irreducible (1PI) diagrams of the model
do not contain $\phi -\varepsilon $ and $\overline{c}-c$ propagators. This
crucial property implies that there is no divergent candidate to renormalize
the term $I_{C}$ of the action. Hence the regularization procedure does not
affect the integrability of the complex structure and subsequently the
metric independence of the action in a structure coordinate neighborhood.

\section{RENORMALIZABILITY}

\setcounter{equation}{0}

The present action does not contain any dimensional parameter like the
ordinary gauge field action. Therefore the dimensionality of the
counterterms will be four. It also admits the enhanced Weyl symmetry (\ref%
{a13}), which does not permit the counterterm (\ref{a2}) with the Weyl
tensor. This means that no pure metric dependent action counterterm will be
generated. The action has also the following discrete symmetry

\ \
\begin{equation}
\begin{array}{l}
a)\ \ell ^{\mu }\Leftrightarrow n^{\mu }\quad ,\quad \phi
_{0}\Leftrightarrow \overline{\phi _{\widetilde{0}}}\quad ,\quad \phi
_{1}\Leftrightarrow \overline{\phi _{\widetilde{1}}} \\
b)\ m^{\mu }\Leftrightarrow \overline{m}^{\mu }\quad ,\quad \phi
_{a}\Leftrightarrow \overline{\phi _{a}}\quad \quad \forall \ a%
\end{array}
\label{e1}
\end{equation}%
which assures that the extended Weyl preserving term $\sqrt{-g}\left( \ell
^{\mu }n^{\rho }F_{\!j\mu \rho }\right) \left( m^{\nu }\overline{m}^{\sigma
}F_{\!j\nu \sigma }\right) $\ does not emerge from the renormalization
procedure.

One might think that the integrability condition of the complex structure is
not necessary for the renormalizability of the action. This is not true,
because the mentioned cases do not exhaust all the possible counterterms.
Notice that the action depends on the tetrad and not directly on the metric
of the spacetime. In fact no metric appears in the action of the model,
where the tetrad vectors $(\ell _{\mu },\,n_{\mu },\,m_{\mu },\,\overline{m}%
_{\mu })$ must be treated as four independent vector fields. Therefore there
may be generally covariant tetrad terms invariant relative to the extended
Weyl symmetry and the above discrete symmetry. The following
1-dimensionality forms and their complex conjugate transform as densities
relative to the extended Weyl symmetry.\ \
\begin{equation}
\begin{array}{l}
(\ell n\partial m)\ ,\ (\ell m\partial \ell )\ ,\ (\ell m\partial m) \\
(nm\partial n)\ ,\ (nm\partial m)\ ,\ (m\overline{m}\partial \ell )\ ,\ (m%
\overline{m}\partial n) \\
\end{array}
\label{e2}
\end{equation}%
Where the compact notation $(\ell n\partial m)$\ has already defined and it
denotes $(\ell n\partial m)=(\ell ^{\mu }n^{\nu }-\ell ^{\nu }n^{\mu
})(\partial _{\mu }m_{\nu })$.\ They may be combined to generate invariant
polynomial and/or non polynomial Lagrangian terms. A typical example of a
4-dimensional Lagrangian symmetric term is the following\ \
\begin{equation}
\begin{array}{l}
\int d^{4}x\sqrt{-g}(\ell n\partial m)(\ell n\partial \overline{m})(m%
\overline{m}\partial \ell )(m\overline{m}\partial n) \\
\end{array}
\label{e3}
\end{equation}%
Notice that this term is not affected (annihilated) by the integrability
conditions of the complex structure. Therefore in principle such a
counterterm could be generated. Its exclusion is implied by the following
argument.

The complex structure integrability conditions give the present action the
form (\ref{a5}). It is apparently tetrad independent. This means that there
is a coordinate system where the action is tetrad independent. Therefore any
geometric term (tetrad dependent) cannot be generated as long as the
renormalization procedure does not change the action term (\ref{a12}) which
imposes the complex structure integrability conditions. This is valid, as it
has already been pointed out at the end of the previous section, because
there is no tetrad-tetrad propagator. It implies that there is no one
particle irreducible loop diagram with $\phi $ external lines. Hence the
action term (\ref{a12}) is not affected by the renormalization procedure
neither other terms with $\phi $ factors can emerge.

\section{REGULARIZATION}

\setcounter{equation}{0}

The expansion around the constant light-cone tetrad separates the
4-dimensional spacetime into two different 2-dimensional spaces, because in
the convenient gauge condition all the field propagators become the product
of two 2-dimensional propagators or one 2-dimensional propagator and a
2-dimensional delta function. This is the characteristic property of the
special gauge condition which is responsible for the finiteness of the loop
diagrams computed below. Any loop-integral turns out to become the product
of two independent 2-dimensional integrals. Therefore the dimensional
regularization must be simultaneously performed in both 2-dimensional
subspaces. It is done by extending the dimension of the $(L_{\mu },\ N_{\mu
})$-subspace into $2\omega $ and the dimension of the $(M_{\mu },\ \overline{%
M}_{\mu })$-subspace into $2\omega ^{\prime }$.

When the dimension of the spacetime changes into $2(\omega +\omega ^{\prime
})$ the number of tetrads changes too. Therefore we first make the
substitutions $2(Lk)(Nk)=k^{2}$ and $2(Mk^{\prime })(\overline{M}k^{\prime
})=k^{\prime 2}$ in all the integrals and after they are dimensionally
regularized. The results are finally contracted with the remaining tetrads
using the formula%
\begin{equation}
E_{a}^{\mu }E_{b}^{\nu }\eta _{\mu \nu }=\eta _{ab}  \label{f1}
\end{equation}%
which does not contain the spacetime dimension. It does appear after the
additional contraction with $\eta ^{ab}$.

The formula of the dimensional regularization, which will be applied are the
called \textquotedblleft 't Hooft-Veltman conjecture\textquotedblright \cite%
{tH-V1972}%
\begin{equation}
\int \frac{d^{2\omega }k}{(2\pi )^{2\omega }}(k^{2})^{\beta -1}=0\qquad
\forall \beta =0,1,2,...  \label{f2}
\end{equation}%
and the following logarithmically divergent 2-dimensional integral%
\begin{equation}
\begin{array}{l}
I_{\rho \nu }=\int \frac{d^{2\omega }k}{(2\pi )^{2\omega }}\frac{k_{\rho
}k_{\nu }}{k^{2}(k-p)^{2}}=\eta _{\rho \nu }\frac{\Gamma (1-\omega )}{2(4\pi
)^{\omega }}\int_{0}^{1}dx[x(1-x)p^{2}+\mu ^{2}]^{\omega -1}+ \\
\qquad +p_{\rho }p_{\nu }\frac{\Gamma (2-\omega )}{(4\pi )^{\omega }}%
\int_{0}^{1}dxx^{2}[x(1-x)p^{2}+\mu ^{2}]^{\omega -2} \\
\end{array}
\label{f3}
\end{equation}%
where the ordinary mass term $\mu ^{2}$ has been introduced in order to
distinguish the ultraviolet from the infrared divergencies. Notice that in
the infrared limit $(\mu ^{2}=0)$ the annihilation of the tadpole diagram ($%
\beta =0$ in the 't Hooft-Veltman conjecture) is rederived\cite{LEIB1975}.

In the present 2-dimensional case $(\omega =1)$ the second term of $I_{\rho
\nu }$ has no ultraviolet divergence, therefore the following integrals,
which appear in the calculations, are finite.%
\begin{equation}
\begin{array}{l}
\int \frac{d^{2}k}{(2\pi )^{2}}\frac{(Lk)}{(Nk)(L\cdot (k-p))(N\cdot (k-p))}%
=i(Lp)^{2}\int_{0}^{1}dx\frac{x^{2}}{x(1-x)(-p^{2})+\mu ^{2}} \\
\\
\int \frac{d^{2}k^{\prime }}{(2\pi )^{2}}\frac{(Mk^{\prime })}{(\overline{M}%
k^{\prime })(M\cdot (k^{\prime }-p^{\prime }))(\overline{M}\cdot (k^{\prime
}-p^{\prime }))}=(Mp)^{2}\int_{0}^{1}dx\frac{x^{2}}{x(1-x)(p^{\prime 2})+\mu
^{2}} \\
\end{array}
\label{f4}
\end{equation}%
where no-primed $k,p$ denote the $(L_{\mu },\ N_{\mu })$-subspace and the
primed $k^{\prime },\ p^{\prime }$ denote the $(M_{\mu },\ \overline{M}_{\mu
})$-subspace components of the 4-momenta $k,\ p$. Analogous results are
found in the $(N^{\mu }N^{\nu }I_{\mu \nu })$ and $(\overline{M}^{\mu }%
\overline{M}^{\nu }I_{\mu \nu })$ contractions.

\section{FIRST\ ORDER\ ONE-LOOP DIAGRAMS\ ARE\ FINITE}

\setcounter{equation}{0}

It has already been stated that there are no loop diagrams with $\phi _{a}(x)
$ external lines. The three possible cases of first order one-loop diagrams
are a) with external tetrads and b) with external gauge fields. I find more
convenient to use the Bogolioubov-Chirkov procedure\cite{B-C1960} for the
computation of the S-matrix one-loop terms as time-ordered products. Only
the main points will be outlined, because it is practically impossible to
present all the calculations here.

\textit{a) Diagrams with two external tetrads. }These diagrams come from the
contractions between internal couplings of $I_{G},$ $I_{C}$\ and $I_{FP}$\
separately. The ghost field contractions give%
\begin{equation}
\begin{array}{l}
\lbrack 2\ ext.\ tetrads\ from\ I_{FP}]=-\gamma ^{2}\int
d^{4}y_{1}d^{4}y_{2}\{:(L\varepsilon _{1}(1))(M\varepsilon _{0}(2)): \\
\cdot \langle T\overline{c}_{1}(1)M^{\mu }\partial _{\mu }(L_{\nu }c^{\nu
}(2))\rangle \langle TL^{\mu }\partial _{\mu }(M_{\nu }c^{\nu }(1))\overline{%
c}_{3}(2)\rangle + \\
:(L_{\rho }\varepsilon _{\widetilde{1}}^{\rho }(1))(\overline{M}_{\tau
}\varepsilon _{0}^{\tau }(2)):\langle T\overline{c}_{1}(1)\overline{M}^{\mu
}\partial _{\mu }(L_{\nu }c^{\nu }(2))\rangle \langle TL^{\mu }\partial
_{\mu }(\overline{M}_{\nu }c^{\nu }(1))\overline{c}_{4}(2)\rangle + \\
:(N_{\rho }\varepsilon _{1}^{\rho }(1))(M_{\tau }\varepsilon _{\widetilde{0}%
}^{\tau }(2)):\langle T\overline{c}_{2}(1)M^{\mu }\partial _{\mu }(N_{\nu
}c^{\nu }(2))\rangle \langle TN^{\mu }\partial _{\mu }(M_{\nu }c^{\nu }(1))%
\overline{c}_{3}(2)\rangle + \\
:(N_{\rho }\varepsilon _{\widetilde{1}}^{\rho }(1))(\overline{M}_{\tau
}\varepsilon _{\widetilde{0}}^{\tau }(2)):\langle T\overline{c}_{2}(1)%
\overline{M}^{\mu }\partial _{\mu }(N_{\nu }c^{\nu }(2))\rangle \langle
TN^{\mu }\partial _{\mu }(\overline{M}_{\nu }c^{\nu }(1))\overline{c}%
_{4}(2)\rangle \} \\
\end{array}
\label{g1}
\end{equation}%
where $:....:$\ denotes the Wick product and the integration variables $%
y_{1},\ y_{2}$ are briefly denoted $1$\ and $2$\ respectively.

After the substitution of the propagators and some well known changes of
variables, (\ref{g1}) takes the following form%
\begin{equation}
\begin{array}{l}
\lbrack 2\ ext.\ tetrads\ from\ I_{FP}]=-\gamma ^{2}\int
d^{4}y_{1}d^{4}y_{2}\{:(L_{\rho }\varepsilon _{1}^{\rho }(1))(M_{\tau
}\varepsilon _{0}^{\tau }(2)): \\
\cdot \left[ \int \frac{d^{2}k}{(2\pi )^{2}}\frac{(L(p-k))}{(Lk)}\right] %
\left[ \int \frac{d^{2}k^{\prime }}{(2\pi )^{2}}\frac{(M(p-k^{\prime }))}{%
(Mk^{\prime })}\right] + \\
+:(L\varepsilon _{\widetilde{1}})(\overline{M}\varepsilon _{0}):\left[ \int
\frac{d^{2}k}{(2\pi )^{2}}\frac{(L(p-k))}{(Lk)}\right] \left[ \int \frac{%
d^{2}k^{\prime }}{(2\pi )^{2}}\frac{(\overline{M}(p-k^{\prime }))}{(%
\overline{M}k^{\prime })}\right] + \\
+:(N\varepsilon _{1})(M\varepsilon _{\widetilde{0}}):\left[ \int \frac{d^{2}k%
}{(2\pi )^{2}}\frac{(N(p-k))}{(Nk)}\right] \left[ \int \frac{d^{2}k^{\prime }%
}{(2\pi )^{2}}\frac{(M(p-k^{\prime }))}{(Mk^{\prime })}\right] + \\
+:(N\varepsilon _{\widetilde{1}})(\overline{M}\varepsilon _{\widetilde{0}}):%
\left[ \int \frac{d^{2}k}{(2\pi )^{2}}\frac{(N(p-k))}{(Nk)}\right] \left[
\int \frac{d^{2}k^{\prime }}{(2\pi )^{2}}\frac{(\overline{M}(p-k^{\prime }))%
}{(\overline{M}k^{\prime })}\right] \} \\
\end{array}
\label{g2}
\end{equation}%
where the defined light-cone notation is used.

Using the formulas of the regularization subsection one can show that all
the above integrals vanish in the context of the dimensional regularization.

The integrals generated by the $I_{C}$\ couplings are analogous to the
previous ones and they are found to vanish too. The expression is too long
to be written down here, therefore only the diagram with $(L\varepsilon
_{1})\ (N\varepsilon _{0})$\ external lines will be presented in order to be
shown how they look like.%
\begin{equation}
\begin{array}{l}
\lbrack (L\varepsilon _{1})\ (N\varepsilon _{0})\ from\ I_{H}]=-\gamma
^{2}\int d^{4}y_{1}d^{4}y_{2}:(L\varepsilon _{1})(N\varepsilon _{0}):\cdot
\\
\cdot \langle T\overline{\phi _{0}}N^{\mu }\partial _{\mu }(\overline{M}%
\varepsilon _{\widetilde{0}}))\rangle \langle TL^{\nu }\partial _{\nu }(%
\overline{M}\varepsilon _{0})\phi _{\widetilde{0}}\rangle = \\
=-\gamma ^{2}\int d^{4}y_{1}d^{4}y_{2}\{:(L\varepsilon _{1})(N\varepsilon
_{0}):\int \frac{d^{4}p}{(2\pi )^{4}}e^{ip(y_{2}-y_{1})}\cdot  \\
\cdot \left[ \int \frac{d^{2}k}{(2\pi )^{2}}(Nk)(L(p-k))\right] \left[ \int
\frac{d^{2}k^{\prime }}{(2\pi )^{2}}\frac{1}{(\overline{M}k^{\prime })(%
\overline{M}(p-k^{\prime }))}\right] \}%
\end{array}
\label{g3}
\end{equation}%
This term vanishes because of the 't Hooft-Veltman conjecture applied to the
$k$-integration.

The diagrams from the $I_{G}$\ couplings, with gauge field contractions, are%
\begin{equation}
\begin{array}{l}
\lbrack 2\ ext.\ tetrads\ from\ I_{G}]=-\frac{\gamma ^{2}}{2}\int
d^{4}y_{1}d^{4}y_{2}\{:(\overline{M}\varepsilon _{0})(\overline{M}%
\varepsilon _{0}):\cdot  \\
\cdot \lbrack \langle T\left( LM\partial A_{j}\right) \left( LM\partial
A_{k}\right) \rangle \langle T\left( M\overline{M}\partial A_{j}\right)
\left( M\overline{M}\partial A_{k}\right) \rangle + \\
+\langle T\left( LM\partial A_{j}\right) \left( M\overline{M}\partial
A_{k}\right) \rangle \langle T\left( M\overline{M}\partial A_{j}\right)
\left( LM\partial A_{k}\right) \rangle ]- \\
-2:(\overline{M}\varepsilon _{0})(N\varepsilon _{1}):[\langle T\left(
LM\partial A_{j}\right) \left( LM\partial A_{k}\right) \rangle \langle
T\left( M\overline{M}\partial A_{j}\right) \left( LN\partial A_{k}\right)
\rangle + \\
+\langle T\left( LM\partial A_{j}\right) \left( LN\partial A_{k}\right)
\rangle \langle T\left( M\overline{M}\partial A_{j}\right) \left( LM\partial
A_{k}\right) \rangle ]- \\
-2:(M\varepsilon _{0})(N\varepsilon _{\widetilde{0}}):[\langle T\left(
LM\partial A_{j}\right) \left( M\overline{M}\partial A_{k}\right) \rangle
\langle T\left( M\overline{M}\partial A_{j}\right) \left( N\overline{M}%
\partial A_{k}\right) \rangle + \\
+\langle T\left( LM\partial A_{j}\right) \left( N\overline{M}\partial
A_{k}\right) \rangle \langle T\left( M\overline{M}\partial A_{j}\right)
\left( M\overline{M}\partial A_{k}\right) \rangle ]+similar\ terms\} \\
\end{array}
\label{g4}
\end{equation}%
This expression is also too long to be written down. I computed all these
integrals and I found that they vanish. The conclusion is that there is no
counterterm with two external tetrads.

\textit{b) Diagrams with external gauge fields.} The number of these
diagrams is quite large, but they can be grouped using the discrete
symmetries (\ref{e1}) of the action. The diagrams with $(LA_{i})(LA_{j})$\
external terms give%
\begin{equation}
\begin{array}{l}
\lbrack ext(LA_{i})(LA_{j})]=-\frac{\gamma ^{2}}{2}\int
d^{4}y_{1}d^{4}y_{2}\ f_{j_{1}i_{1}k_{1}}\
f_{j_{2}i_{2}k_{2}}:(LA_{i_{1}})(LA_{i_{2}}):\cdot  \\
\qquad \qquad \cdot \lbrack \langle T\left( MA_{k_{1}}\right) \left(
MA_{k_{2}}\right) \rangle \langle T\left( N\overline{M}\partial
A_{j_{1}}\right) \left( N\overline{M}\partial A_{j_{2}}\right) \rangle + \\
\qquad \qquad +\langle T\left( MA_{k_{1}}\right) \left( N\overline{M}%
\partial A_{j_{2}}\right) \rangle \langle T\left( N\overline{M}\partial
A_{j_{1}}\right) \left( MA_{k_{2}}\right) \rangle + \\
\qquad \qquad +\langle T\left( MA_{k_{1}}\right) \left( \overline{M}%
A_{k_{2}}\right) \rangle \langle T\left( N\overline{M}\partial
A_{j_{1}}\right) \left( NM\partial A_{j_{2}}\right) \rangle + \\
\qquad \qquad +\langle T\left( MA_{k_{1}}\right) \left( NM\partial
A_{j_{2}}\right) \rangle \langle T\left( N\overline{M}\partial
A_{j_{1}}\right) \left( \overline{M}A_{k_{2}}\right) \rangle +c.c.]= \\
\\
\qquad =-\frac{i\gamma ^{2}C}{16(4\pi )^{2}}\int d^{4}y_{1}d^{4}y_{2}\ \
\int \frac{d^{4}p}{(2\pi )^{4}}e^{ip(y_{2}-y_{1})}:(LA_{i_{1}})(LA_{i_{2}}):%
\cdot  \\
\qquad \cdot \delta _{i_{1}i_{2}}\ (Np)^{2}(Mp)^{2}(\overline{M}%
p)^{2}I_{1}(p^{\prime \prime 2})[2I_{2}(-p^{\prime 2})-I_{1}(-p^{\prime 2})]
\\
\end{array}
\label{g6}
\end{equation}%
where $p^{\prime 2}=(p^{0})^{2}-(p^{1})^{2}$ ,  $p^{\prime \prime
2}=(p^{2})^{2}+(p^{3})^{2}$,\  $f_{jik}\ f_{j^{\prime }ik}=C\delta
_{jj^{\prime }}$ and the final integrals%
\begin{equation}
I_{r}(k^{2})=\int_{0}^{1}dx\frac{x^{r}}{x(1-x)k^{2}+\mu ^{2}}  \label{g7}
\end{equation}%
are finite. Hence the diagrams with $(LA_{i})(LA_{j})$\ external terms are
finite. All the other diagrams with two gauge field lines vanish or they are
finite like the above. On the other hand the one-loop diagrams with two
external gauge field lines and internal ghost lines vanish too because of
the $k$-integration. Hence my conclusion is that there is no first order
one-loop counterterms with two external gauge fields.

In order to see whether the gauge field coupling constant is renormalized
one has to study the second order one-loop diagrams. All these diagrams with
three external gauge fields and with two and three internal gauge fields
have been written down. Their number is quite large, but they can be grouped
using the above discrete symmetry. Investigating these diagrams, I found
that they are all finite to. This implies that there is no first order
coupling constant renormalization, which means that the first term of the
function $\beta (\gamma )$ of the renormalization group equation vanishes.

\section{DISCUSSION}

\setcounter{equation}{0}

For a long time, it was believed that there is no four dimensional
renormalizable generally covariant Lagrangian model without higher order
derivatives. Recall that this is one of the arguments which turned Physics
research to strings. The present model shows that this belief is not true.
The remarkable point is that the renormalizability is achieved through the
standard technique of enhancing the symmetry. In the present model the Weyl
(conformal) symmetry over the metric is extended to every vector of the
integrable (null) tetrad $(\ell _{\mu },\,n_{\mu },\,m_{\mu },\,\overline{m}%
_{\mu })$. This symmetry was achieved after a slight modification of the
Yang\_Mills action. But this modification has the following severe
consequences. The action is no longer metric dependent. Instead it is only
complex structure dependent like the two dimensional string action. Besides
this extended Weyl symmetry does not seem to permit the introduction of
fermionic fields. All my efforts to find symmetric fermionic action terms
have failed. This feature may not cause a problem to the physical content of
the model, because some geometric solitons of the model have fermionic
gyromagnetic ratio. This means that the present model may not be
supersymmetrizable and may not even need a supersymmetrization to include
fermions. The other essential difference between the present action and the
ordinary Yang Mills action is at the generated static potential. The present
action generates a confining linear static potential instead of the well
known $\frac{1}{r}$\ Coulomb potential of the ordinary Yang-Mills action.
This means that the expected "quark confinement" is now perturbative without
any reference to the not yet proved "infrared confinement" of ordinary
gluonic action.

The present proof of the renormalizability of the model is based on the
exclusion of all possible counterterms. It would be interesting to prove it
using the conventional method of Ward identities, which are very complicated
in the present gauge conditions. The first loop diagrams confirmed the
renormalizability of the model and they indicate that it may be finite. In
any case, it would be interesting to compute the loop diagrams with four
external tetrads, which could generate the symmetric term (\ref{e3}).
Finiteness of these diagrams would persuade us that the present model is
something special.

It is well known that anomalies could destroy renormalizability. The
finiteness of the first loop diagrams implies no first order anomalies, but
they cannot be excluded to appear in higher orders.

In current terminology, a Lagrangian model is called finite if all its
transition amplitudes on mass shell are finite without making use of any
infinite renormalization either of the field or of the coupling constants.
These amplitudes (on mass shell) do not depend on the regularization
procedure or the imposed gauge condition, therefore their finiteness should
not depend on these two choices either. The general Green functions of a
finite field theoretical model may diverge, depending on the used gauge
conditions. Apparently the existence of a gauge condition, which makes the
Green functions finite, imply finiteness of the model. This formal reasoning
works well in the case of supersymmetric Yang-Mills (SYM) model. It has been
conjectured that the four dimensional $N=4$ supersymmetric Yang-Mills (SYM)
model is finite\cite{MAND1983}, while the six and ten dimensional SYM models
are not finite\cite{RAG1983}. Therefore the fact that in the precise
convenient gauge, which was used in the present calculations, the Green
functions are finite, implies that the present model is also finite in the
first order approximation. In a different gauge condition (e.g. Landau or
Feynman) the Green functions may not be finite but the cross-sections must
be finite.

\bigskip

\end{document}